\documentclass[twocolumn,showpacs,preprintnumbers,amsmath,amssymb]{revtex4}
\usepackage{graphicx}
\usepackage{dcolumn}
\usepackage{bm}
\begin{document}
\title{Exotic hadron production  in quark combination model}

\author{Wei Han$^1$, Shi-Yuan Li$^2$, Yong-Hui Shang$^2$,
Feng-Lan Shao$^1$, Tao Yao$^2$}

\affiliation{$^1$Department of Physics, Qufu Normal University,
Qufu, 273165, P. R. China\\ $^2$Department of Physics, Shandong
University, Jinan, 250100, P. R. China}

\date{\today}

\begin{abstract}
 The philosophy on production of exotic hadrons
(multi-quark states) in the framework of quark combination model is
investigated, taking $f_0(980)$  as an example. The production rate
and $p_T$ spectra of $f_0(980)$ considered as $(s\bar s)$ or $(s\bar
q\bar s q)$ respectively, are calculated and compared in Au+Au collisions at
$\sqrt{s_{NN}}=200GeV$. The unitarity of
various combination models, when open for exotic hadron production,
is addressed.
\end{abstract}



\maketitle

\section{Introduction}

 The basic quanta of Quantum ChromoDynamics (QCD),
quarks and gluons, are  confined in their bound states, hadrons. In
any high energy interaction,
 the produced colour-singlet (CS) (anti)quark system eventually
transits to various hadron states with the total probability exactly
1:
\begin{equation}
\label{ueq1} \sum_h |<h|U|q>|^2=<q|U^+U|q>=1.
\end{equation}
Here we introduce the unitary time-evolution operator $U$ to
describe the hadronization process. For the quark state
 $|q>$ and the corresponding hadron state $|h>$, the
matrix element $U_{hq}=<h|U|q>$ describes the transition amplitude.
$U_{hq}$ is determined by QCD but beyond the present  approach of
calculation. This leaves the space for various hadronization models
to mimic this transition process.  As a matter of  fact from
experiments,
\begin{equation}
\label{ueqless1} \sum_{h=B,\bar B,M} |<h|U|q>|^2 \sim 1-\varepsilon,
~ \varepsilon \rightarrow 0^+,
\end{equation}
here $B, \bar B, M$ denote baryon, antibaryon and meson respectively.

 Na\"{i}vely from the group theory, {\it colour
confinement} seems not so restrict as  Eq. (\ref{ueqless1}). The CS
state, i.e., the invariant, totally antisymmetric representation of
the $SU_C(3)$ group, requires at least one quark and one antiquark,
or three (anti)quarks (These are just the constituent/valence quark
numbers for mesons and baryons, respectively), but more (anti)quarks
can also construct this
 representation, hence possibly to form a CS ``hadron''. To
name some possibilities, two quark-antiquark pairs, a
quark-antiquark pair with 3 (anti)quarks, six (anti)quarks, etc.,
are to be called exotic hadrons in this paper  \footnote{There is
also, the possible existence of bound states including
 gluons, which is not covered in this paper and the name exotic
hadron here does not include glueball or hybrid
except explicit statements.}.
 Until now, no experiment can definitely show    $\varepsilon$ in
Eq. (\ref{ueqless1}) is exactly 0 or a small but {\it non}-vanishing
number. If definitely $\varepsilon=0$, there must be underlying
properties of QCD which we still not very familiar. Even
$\varepsilon$ is not vanishing, its smallness, definitely confirmed
by experiments and shown in Eq. (\ref{ueqless1}),  also provides
interesting challenges, especially on hadronization models. The
small production rate of a special kind of exotic hadron seems easy
to be adopted. However, taking into account so many possibilities to
construct the CS representations by {\it various} numbers of
(anti)quarks, that the total sum of them is still quite small, is
very nontrivial as property of QCD and even nontrivial for a
hadronization model to reproduce.

More concretely,  we investigate the quark combination models
\cite{anisovich, bjorken} when open for the production of
multi-quark states.  At first sight, the quark combination model is
the most feasible in calculation, to just allow the desirable number
of (anti)quarks to combine together. But to get the total production
probability of all the exotic hadrons fulfilling  Eq.
(\ref{ueqless1}) with a universal combination rule is nontrivial. In
this paper, we investigate the production of exotic hadrons via a
quark combination model proposed by the Shandong Group (SDQCM)
\cite{xie1,xie2,xie3,xie4,xie5,xie6}. We demonstrate that to treat
the {\it production process} of  all multi-quark states as hadron
molecule formation could be a practical and self-consistent way,
respecting Eqs. (\ref{ueq1}, \ref{ueqless1}),  the unitarity of
hadronization models (see also \cite{yangcb}, and refs. therein).

  Of all
the ``on market'' combination models, SDQCM  is unique for its
combination rule  wisely designed, so that mesons and baryons
exhaust the probability of all the fates of the (anti)quarks in a CS
system. It has been realized in Monte-Carlo  programmes and
tested against data from experiments in high energy $e^+e^-$
annihilation and $pp$ collisions, and recently it has been
successfully extrapolated to ultra-relativistic heavy ion
collisions \cite{shao1,shao2,yao1}, reflecting the universality
of the hadronization mechanism.  Most recently, application of the
combination rule to the open charm and bottom sector for
Relativistic Heavy Ion Collider (RHIC) experiments, without any
more parameters for the hadronization, further demonstrates its validity
and provides  opportunities against more critical tests \cite{yao2}.

The main idea is to line up $N_{q}$ (anti)quarks in a
one-dimensional order in phase space, e.g., in rapidity, and then
let them combine into initial hadrons one by one following a
combination rule \cite{xie1,xie2,xie3,xie4,xie5,xie6}, three (anti)quarks
or a quark-antiquark pair in the neighborhood  can form a (anti)baryon or
a meson, respectively. The cases of
more numbers ($>3$) of (anti)quarks to combine into hadrons
automatically disappear. The
inclusive cross section is proportional to the product of the quark
number densities, leading to $B/M$ enhancement in a region (which is
in fact one of the motivations for the burst of combination models
at RHIC). Without this special combination rule, more-quark state could be
even enhanced \cite{yangcb}. This will conclude a nonsense result
that a system of (anti)quarks could combine into a giant ``quark
ball'', with large number of constituent quarks. So, one of the key
properties of combination models is  whether a model can shut off
the possibility that exotic number of (anti)quarks combined into
cluster with a large probability.

To introduce the small probability $\varepsilon$ in Eq.
(\ref{ueqless1}) to SDQCM, similar as the framework to calculate the
resonance production of  heavy quark bound states \cite{changetal}, we consider the
exotic hadron production as hadron molecule production and project
the free mesons and/or baryons (anti-baryons) states onto their
bound states in terms of the wave functions.  In this paper we
first clarify that  it does {\it not} mean that, in the bound state,
quarks from each hadron  should  keep in CS respectively,
but the colour interactions
can transit the whole  molecule  into a ``real'' exotic hadron by some
probability, which will be introduced in the following. This is
especially reasonable for the circumstance like RHIC, where the
large number of hadrons and large bulk of thermalized  area allow
interactions between hadrons lasting for a long time.

In the following we first discuss the colour state and
``definition'' of quark number of the exotic hadrons (Sec. II).
Then, taking $f_0(980)$  as a working example, we describe the
calculations of the production of the exotic hadrons within the
quark combination model and discuss the results (Sec.III).
Sec. IV is the conclusion.

\section{Colour state and quark number of the exotic hadrons}

\subsection{Colour state in an exotic hadron}

All kinds of Exotic hadrons have one common property, which is that
the (anti)quarks can be grouped into several clusters, with each
cluster {\it possibly} in CS. Hence, could the exotic hadrons just
be meson and/or baryon molecules?  However, the ways of grouping
these (anti)quarks are not unique, as it is simply known from group
theory that the reduction ways for a direct product of several
representations are not unique.  Furthermore, these clusters need
not necessarily be in CS respectively, since only requirement is the
whole set of these clusters in CS. For example, the system $q_1
\bar{q_2} q_3 \bar{q_4}$ (the constituents of a ``tetraquark'') can
be decomposed/clustered in the following ways:
\begin{eqnarray}
(q_1 q_3)_{\bar 3} \otimes (\bar{q_2}
\bar{q_4})_{3} \rightarrow 1\\
(q_1 \bar q_2)_{1~ or ~8} \otimes (q_3
\bar q_4)_{1~ or ~8} \rightarrow 1\\
\cdot \cdot \cdot \nonumber
\end{eqnarray}
Here we just mention that such group theory analysis is applicable to the 
quark states  as well as the quark field operators \cite{pire}. 
In the above example, only  the second case, when these two $q\bar
q$ pairs  are in CS respectively, it seems possible to be considered
as a hadron molecule. But dynamically, the colour interactions in
the system via exchanging gluons can change the colour state of each
separate cluster, so each kind of grouping/reduction way seems no special physical
meaning. Such an ambiguity, which has been considered in many
 hadronization and decay processes as ``colour
recombination/rearrangement'' \cite{our1, our2, gsj}, obstacles the possibility
to consider the exotic hadron in a unique and uniform way, while
leads to the possibility of introducing some phenomenological
duality. Namely, even we consider the production of exotic hadron as
``hadron molecule'' formation, the subsequent colour interactions
 in the system can eventually transit  this
``molecule'' into a ``real''  exotic hadron, at least by some
probability.


 \subsection{How to count the quark number in an exotic hadron}

Some kinds of exotic hadrons have exotic quantum number(s), e.g.,
one kind pentaquark ($qqq\bar{s}q$), has $+1$ baryon number but $+1$
strange number. If a hadron with such quantum numbers is
experimentally confirmed, one seems to have to introduce five
valence (anti)quarks. However, in many other cases, there exist
parallel explanations because of non-exotic quantum number of a
certain hadron. The tetraquark or four-quark state discussed in this
paper is an example. One of the  candidates is $f_0(980)$. It
is considered as an orbit-excited $l=1$ regular meson ($s \bar s$)
\cite{f02}, but argued possibly to be a four-quark state ($s \bar q
\bar sq$) by others \cite{f04}. So one naturally raises the question
relating with this ambiguity, how on earth can one count the number
of quarks in a hadron?  Even, what is the meaning of ``number of
quarks in a hadron''?  This may be one of the most ambiguities in
the physical picture of the exotic hadrons, because of  lack of
complete understanding of the confinement property of QCD. It is not
clear what the quantum field theory definition of the ``constituent
quark'' is. As a consequence, it is ambiguous how to ``count'' the
quark number in a hadron. Here we just state the different pictures
of a proton, one is  parton model bursting from the deeply inelastic
scattering and other high energy interaction processes, the other is
the ``static'' quark model, corresponding to the properties of a
proton at rest. There is no satisfactory, especially quantitatively
relation between these two pictures. Even the consideration of
higher Fock states or pair excitations \cite{excitation} can not
remove the gap.

It is well-known that factorization theorem confirms the parton
fragmentation picture to describe the hadron production \cite{fac}.
But this partonic picture is only valid in inclusive processes with
hard interactions involved. For the low $p_T$ particles, e.g, those
from ``fragmentation'' of the hadron remnant in hadron-hadron
interactions, or in a very complex circumstance like in heavy ion
collisions, this picture faces both challenges from theoretical as
well as experimental aspects. It is difficult to prove the
factorization theorem for these complex cases. Furthermore,   the RHIC data expose
some properties difficult to be understood from fragmentation
picture, such as the high  $p/\pi$ ratio at intermediate transverse
momenta \cite{p/pi} and the quark number scaling of hadron elliptic
flows \cite{elliptic}, but these can be explained by coalescence or
(re)combination models \cite{recombination,coalescence}. Combination
models include
 the following picture: 1) the production of quarks,
which is considered as the ``dressed'' quarks, i.e.,
constituent/valence quarks; 2) these quarks combined to certain
hadrons, e.g., $q \bar q$ combined to meson. Within these models,
the quark number in a hadron has a model-dependent but clear
``definition'', i.e., ``the number of (anti)quarks'' of a certain
hadron is the number of (anti)quarks involved in combining into the
hadron in the {\it production process}. So, as suggested by
\cite{recombination}, one can ``count'' this number experimentally
by measuring the $v_2$ in  heavy-ion collisions at RHIC, since $v_2$
is proportional to the number of quarks combined to the certain
hadron, and the following inner interactions between (anti)quarks in
the hadron will not change this property of global movement. This
makes the quark number an ``observable'', by extrapolating the
combination picture to open for all kinds of exotic hadrons.
Treating exotic hadron production as hadron molecule formation will
not change this fact that $v_2$ is still proportional to the total
quark number in the exotic hadron, which is a straightforward result
of associative law of addition.

\section{the production of $f_0(980)$ at RHIC}

Now we  apply the above discussions into  an example, $f_0(980)$.
 This particle is considered  as a tetraquark  with the flavour
 content $s\bar q\bar s q$ \cite{maiani}.   The traditional consideration for it as
  $l=1$ state of $s \bar s$,
has been included in  SDQCM with all the other $l=1$ hadrons in the
most recent version \cite{yao3}. We also calculate that case for
comparison.

As we mentioned above, the conventional SDQCM can calculate the free
meson and baryon distributions, e.g., two meson distributions
 $\frac{E_1E_2d\sigma}{d^3p_1 d^3p_2}$.
Projecting onto the meson molecule state, we can get the bound state
distribution:
\begin{eqnarray}
 \frac{Ed\sigma^{N}}{d^3p} &= &\sum_{N_1N_2}\int \frac{d^3p_1}{E_1}
 \frac{d^3p_2}{E_2} 
\Big |<p_1,p_2,N_1,N_2|p,N> \Big |^2 \nonumber\\
& \times &\frac{E_1E_2d\sigma^{N_1,N_2}}{d^3p_1 d^3p_2}.
 \label{xsec}
\end{eqnarray}
In the above equation, the $N, N_1, N_2$ refer to discrete quantum
numbers, corresponding to the meson molecule state and the two
mesons, respectively. The conservation relations like $ \delta^4
(p-p_1-p_2)$ and $\delta_{N_1\otimes N_2, N}$ are indicated in the
projection of the state vectors. In order to get this factorized
formulae, some interference terms are missed. Since the free hadron
distribution is calculated by the Monte-Carlo programmes, which can not give
the amplitude but only its square. And only this factorized form can
be used in the  Monte-Carlo programmes.

The projection of the discrete quantum numbers such as flavor,
isospin, angular momentum, charge and space parity etc., are easy to
be calculated. In the case of angular momentum projecting, spin
counting is assumed. Though the system is not assumed to be
non-relativistic, only the lowest possible orbit angular momentum is
considered (i.e., Since in generally, the Parity requires odd or
even $l$ value, in the case of even $l$, we only consider $l=0$,
while for odd, only $l=1$ ).
The detailed analysis on the projection of discrete quantum numbers
is outlined in Table \ref{corr}.

The phase space wave function of the exotic hadron in terms of
mesons and/or baryons is not definitely available. One of the
reasons is that, as  argued above,   the subsequent colour
interactions in the ``molecule''  ruin a unique structure to be
described by definite wave function. To mimic the combination
process, we use the same physical picture as quarks combining into
regular hadrons in our model, i.e., near rapidity correlation \cite{xie1}. This
says that only two mesons in the neighborhood on the rapidity axis
have the chance to combine together. Contrary to the case of quarks,
in which confinement property requires all quarks must be combined
into certain hadrons, the hadrons need not necessarily to be
combined to some molecules. Rather, they may have large probability
to be free. It is natural to introduce a parameter $x$, smaller than
1, to parameterize the probability for two mesons neighbored
 to combine into a bound state. So, $x$ not only reflects the
information of some specific exotic hadrons, but also reflects the
interactions among the  meson and (anti)baryon system.  This assures
the total results  respecting  the fact  that all the quark states and
all the hadron states (mesons, (anti)baryons and exotic hadrons)
respectively form two complete sets of bases for the same Hilbert
space. This unitary transformation assures the unitarity of the
quark combination model.

Unfortunately, the unknown wave functions of all kinds of exotic
hadrons mean $x$ is almost a free parameter, only constrained by the
data of mesons and baryons, as well as how many kinds of exotic
hadrons to be considered.  It is clear that to an extreme if we have infinite
kinds of exotic hadrons, $x$ should be vanishing, expecting infinite
number of vanishing variables (production rates corresponding to  each  certain
 exotic hadron) summing up to get a finite small
result (the total production rate of all exotic hadrons).
For demonstration, here we only consider the existence of four-quark states as
exotic hadrons, but still with two choices: One case is that only
$f_0(980)(s\bar q \bar s q)$ exists in the world, the other is not
only $f_0(980)(s\bar q \bar s q)$ but also any other tetraquark
states (isospin muti-states) to be allowed to produce by our model.
In the latter case more mesons are ``used up'' to get the molecule,
so the $x$ must be smaller
than the former case. We first tune the Monte-Carlo programme to produce the
multiplicities of mesons and baryons at the central value of the
experimental data.  The baryons are not used to  produce the exotic
hadrons, so is definitely determined by experimental data.
Then we tune $x$  so that all the mesons used up should
not exceed the error bar of the experimental data (within 5 per cent
 \footnote{Now we take e.g., error of $\phi$'s data as one of the
most precise. Its production rate $\frac{dN}{dy}$ is $7.70 \pm 0.30$
in midrapidity region from the central Au+Au collisions at
$\sqrt{s_{NN}}=200GeV$ \cite{phi}. Its error is within $5\%$.}). The
results are   in Table \ref{fo} and Fig. 1.  From the above discussion,
it is more fair to take
them as uplimits of the production rates.

\begin{table}

\begin{tabular}{|c|c|c|c|}  \hline
meson pair & $ |<J_1,J_2|J>|^2$ & $|<C_1,C_2|C>|^2$ &
$|<I_1,I_2|I>|^2$
 \\ \hline
$\phi,\omega$ & $1/9$ & 1 & 1   \\ \hline$\eta,\eta$ & 1 & 1 & $4/9$
\\ \hline $\eta,\eta'$ & 1 & 1 & $5/9$
  \\ \hline $\eta',\eta'$ & 1 & 1 & $4/9$   \\ \hline $K^+,K^-$ &
1 & 1 & $1/2$   \\ \hline $K^0,\bar K^0$ & 1 & 1 & $1/2$
 \\
\hline \hline

\end{tabular}
\caption{The projection of discrete quantum numbers (I, C, P)
 of  $M_1, ~M_2$  to $f_0(980)(s\bar q\bar sq)$.
The cases with much smaller values of $|<p_1,p_2,N_1,N_2|p,N>|^2$
  are included in the programme but ignored in the above list.}
\label{corr}
\end{table}

\begin{table}
\begin{tabular}{|c|c|c|c|}
  \hline
   & x & $f_0(980)(s\bar q\bar sq)$ &  $f_0(980)(s\bar s)$ \\

 \hline
$(a)$ & 0.60 &1.63 &  \\
\cline{1-3}
 $(b)$ & 0.24 & 0.65 & \raisebox{1.2ex}[0pt] {\hfill0.68\hfill{}} \\

 \hline
\hline
\end{tabular}
\caption{The production rate of $f_0(980)$ at midrapidity
(within one rapidity unit) for central Au+Au collisions at
$\sqrt{s_{NN}}=200GeV$.  $(a)$ corresponds the
case only one exotic hadron, i.e., $f_0(980)(s\bar q\bar sq)$;  $(b)$
denotes the case  that we consider all the isospin multi-states.}
\label{fo}
\end{table}



Recently, argued in Ref. \cite{maiani}, from the production rate,
one can extract the information of the constituent quark number. The
authors employ the combination model proposed by \cite{f04},
predict and compare the production rate of $f_0(980)$ for two cases: $s \bar s$
or $s\bar q  \bar s q$.  However, whether this idea is
practical  relies on whether it is ``model-independent''.
From our calculations list in table  \ref{fo}, one can see
different results from \cite{maiani}.
 If we take into account the case that many other kinds of exotic hadrons
 could also exist in nature  and formed from combination of hadrons, the
production rate of $f_0(980)(s\bar q\bar sq)$ should be even
smaller. So,  this implies that the production rates depend on the
mechanism of production, and  are  not straightforwardly possible to be
related with the constituent quark number in the framework of various quark
combination models.

For the transverse momentum spectrum (Fig.1), it
is clear the  four quark state is harder, common property of all
combination models. This is the same reason for $B/M$ enhancement in
mid-$p_T$ region. It is quite interesting to point out that, in
principle the $p_T$ spectrum of such kind of exotic hadrons
formed from combination of mesons and/or (anti)baryons can be
fixed in our model, {\it independent} of the {\it quark
distributions}. The free hadron spectra in Eq.(5) can be fixed by
experimental data, which is completely model-independent.
Furthermore, $x$ can only change the absolute value but not the
shape of spectrum. So, if some exotic hadron (maybe not $f_0(980)$)
is produced by combination of regular hadrons at RHIC, we can
predict the shape of its spectrum without any ambiguity.

\begin{figure}
\includegraphics[scale=0.33]{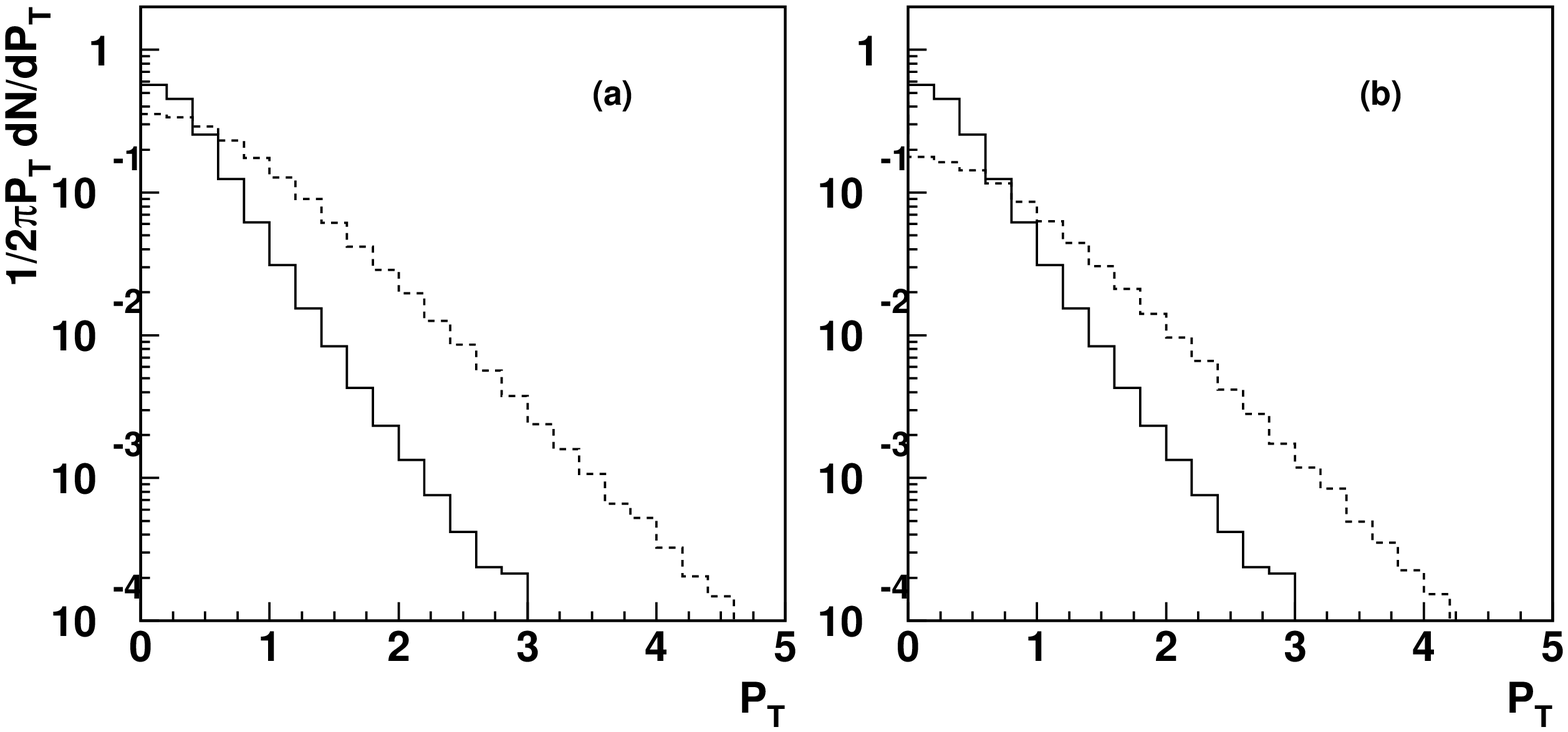}
\caption{Hadron spectra of $f_0(980)$ at midrapidity as a function
of transverse momentum $p_T$ for central Au+Au collisions at
$\sqrt{s_{NN}}=200GeV$. $f_0(980)(s\bar s)$ (solid line) compared
with $f_0(980)(s\bar q \bar sq)$ (Dashed line). (a) and (b)
correspond to $a$ and $b$ in table  \ref{fo} respectively.}

\end{figure}

\section{Conclusion}

In this paper, we propose a way to calculate the production rates of
exotic hadrons (multi-quark states) within the framework of SDQCM.
We point out that this
special combination model employs a specific combination rule to
shut off the possibility that more than three (anti)quarks can
combine into some hadron. The unitarity is automatically kept but
may be too restrict if there could be exotic hadrons, since the
to-date experiments only assure the rareness of exotic hadron
production, as shown in   Eq. (\ref{ueqless1}). By an analysis on
the complexity of the  colour structure of the exotic hadrons, we
suggest that one can introduce the small section  of exotic hadron
production by treating ({\it only}) its production process as hadron
molecule (bound state) formation. Taking the possible four-quark
structure of $f_0(980)$ as an example, we discuss two special cases,
i.e., only one exotic hadron $f_0(980)(s\bar q\bar sq)$ exists in
nature, or  all the isospin multi-states of $(s\bar q\bar sq)$
exist, respectively, and  comparing with the regular  meson
structure of $f_0(980)(s\bar s)$.  This is helpful in the exotic
hadron study in experiments like RHIC.

\acknowledgments This work is partially supported by  National
Natural Science
 Foundation of China
(NSFC) with grant No.10775089 and 10775090. The authors thank the
discussions with members of the particle physics group of Shandong
University, especially Prof. Xie Qu-Bing for his initiation of the
SDQCM as well as numerous encouragements.


\begin{thebibliography}{99}


\vspace{3mm}

\bibitem{anisovich}
V. V. Anisovich and V. M. Shekhter, Nucl. Phys. B {\bf 55}, 455
(1973).
\bibitem{bjorken}
J. D. Bjorken and G. R. Rarrar, Phys. Rev. D {\bf 9}, 1449 (1974).
\bibitem{xie1}
Q.~B.~Xie and X.~M.~Liu , Phys. Rev. D {\bf38}, 2169 (1988).
\bibitem{xie2}
Z.~T.~Liang and Q.~B.~Xie, Phys. Rev. D  {\bf43}, 751 (1991).
\bibitem{xie3}
 Q.~Wang and Q.~B.~Xie, J. Phys. G {\bf21}, 897-904 (1995).
\bibitem{xie4}
 J.~Q.~Zhao, Q.~Wang and Q.~B.~Xie, Sci. Sin. A {\bf38}, 1474-1483 (1995).
\bibitem{xie5}
 Q.~Wang, Z.~G.~Si and Q.~B.~Xie, Int. J. Mod. Phys. A {\bf11}, 5203-5220 (1996).
\bibitem{xie6}
Z.~G.~Si, Q.~B.~Xie and Q.~Wang, Commun. Theor. Phys. {\bf28}, 85-94
(1997).
\bibitem{yangcb}
  see, e.g., C.~B.~Yang, J. Phys. G {\bf 32}, L11 (2006).  
\bibitem{shao1}
F.~L.~Shao, Q.~B.~Xie and Q.~Wang, Phys. Rev. C {\bf 71}, 044903
(2005).
\bibitem{shao2}
 F.L.Shao, T.Yao and Q.B.Xie, Phys.Rev.C {\bf 75},034904 (2007).

\bibitem{yao1}
 T.Yao, Q.B.Xie and F.L.Shao, Chinese. Phis. C {\bf 32},356 (2008). 
\bibitem{yao2}
T.~Yao, W.~Zhou and Q.~B.~Xie, Phys. Rev. C {\bf 78}, 064911,
(2008).
\bibitem{changetal}
C.~H.~Chang,
  Nucl.\ Phys.\  B {\bf 172} (1980) 425;
R.~Baier and R.~Ruckl,
  Z.\ Phys.\  C {\bf 19} (1983) 251;
J.~H.~Kuhn, J.~Kaplan and E.~G.~O.~Safiani,
  Nucl.\ Phys.\  B {\bf 157} (1979) 125.

\bibitem{pire}
I.~V.~Anikin, B.~Pire and O.~V.~Teryaev,
  Phys.\ Lett.\  B {\bf 626}, 86 (2005).

\bibitem{our1}
 W.~Han, S.~Y.~Li, Z.~G.~Si and Z.~J.~Yang, Phys. Lett. B {\bf 642}, 62-67 (2006).

 \bibitem{our2}
 Z.~G.~Si, Q.~Wang and Q.~B.~Xie,
  Phys.\ Lett.\  B {\bf 401} (1997) 107;
  Q.~Wang, Q.~B.~Xie and Z.~G.~Si,
  Phys.\ Lett.\  B {\bf 388} (1996) 346.

\bibitem{gsj}
G.~Gustafson and J.~Hakkinen,
  Z.\ Phys.\  C {\bf 64} (1994) 659;
 T.~Sjostrand and V.~A.~Khoze,
  Z.\ Phys.\  C {\bf 62} (1994) 281.





\bibitem{f02}
A. Deandrea, R. Gatto, G. Nardulli, A. D. Polosa and N. A.
Tornqvist, Phys. Lett. B {\bf 502}, 79 (2001).

\bibitem{f04}
see, e.g., L. Maiani, F. Piccinini, A. D. Polosa and V. Riquer,
Phys. Rev. Lett. {\bf 93}, 212002 (2004); Phys. Rev. D {\bf 71},
014028 (2005); R. L. Jaffe, Phys. Rept. {\bf 409}, 1 (2005) [Nucl.
Phys. Proc. Suppl. {\bf 142}, 343 (2005)].
\bibitem{excitation}
B. S. Zou, D. O. Riska, Phys. Rev. Lett. {\bf 95}, 072001 (2005);
B. C. Liu, B. S. Zou, Phys. Rev. Lett. {\bf 96}, 042002 (2006);
B. C. Liu, B. S. Zou, Phys. Rev. Lett. {\bf 98}, 039102 (2007).

\bibitem{fac}
John C. Collins, Davison E. Soper, George Sterman. Published in
Adv.Ser.Direct. High Energy Phys. {\bf 5}, 1-91 (1988)
[arXiv:hep-ph/0409313].
\bibitem{p/pi}
K. Adcox $et al.$, (PHENIX Collaboration), Phys. Rev. Lett. {\bf
88}, 022301 (2002);
Phys. Rev. Lett, {\bf 88}, 242301(2002);
C.Adlex $et al.$, (STAR Collaboration), Phys. Rev. Lett, {\bf 86},
4778(2001).

\bibitem{elliptic}
 I. Tserruya, Nucl. Phys. A {\bf774}, 415 (2006) [arXiv:nucl-ex/0601036].
\bibitem{recombination}
R. J. Fries, B. M$\ddot{u}$ller, C. Nonaka and S. A. Bass, Phys.
Rev. Lett. 90, 202303 (2003);
Phys. Rev. C {\bf68}, 044902 (2003).
\bibitem{coalescence}
V. Greco, C. M. Ko and P. Levai, Phys. Rev. C {\bf 68}, 034904
(2003);
Phys. Rev. Lett. {\bf 90},  202302 (2003);
R. C. Hwa and C. B. Yang, Phys. Rev. C {\bf 70},  024905 (2004);
Phys. Rev. C {\bf 67}, 034902 (2003).

\bibitem{maiani}
L.~Maiani, A.~D.~Polosa, V.~Riquer and C.~A.~Salgado, Phys. Lett. B
{\bf 645}, 138-145 (2007).

\bibitem{yao3}
T.~Yao, Ph. D Thesis of Shandong University (2009).
\bibitem{phi}
J.~Adams $et al.$, (STAR Collaboration), Phys. Lett. B {\bf 612},
181-189, (2005).



\end{thebibliography}
\end{document}